\documentclass[printer]{aa}  
\usepackage{amsmath}
\usepackage{tablefootnote}
\usepackage{bm}
\usepackage{graphicx}
\usepackage{txfonts}
\usepackage[colorlinks,linkcolor=blue,citecolor=blue,linktocpage=true,breaklinks, plainpages=false,urlcolor=blue]{hyperref}

\bibpunct{(}{)}{;}{a}{}{,}
\RequirePackage{color}

\usepackage{booktabs}
\begin{document} 
\title{\ion{Ti}{I} lines at 2.2\,$\mu$m as probes of the cool parts of sunspots}

\author{H. N. Smitha \inst{1}, J.~S. {Castellanos~Dur\'{a}n} \inst{1}, S. K. Solanki \inst{1,2}, S. K. Tiwari \inst{3,4}}
\institute{Max-Planck-Institut f\"ur Sonnensystemforschung, Justus-von-Liebig-Weg 3, D-37077 G\"ottingen, Germany \and
School of Space Research, Kyung Hee University, Yongin, Gyeonggi, 446-701, Republic of Korea \and
Lockheed Martin Solar and Astrophysics Laboratory, 3251 Hanover Street, Bldg. 252, Palo Alto, CA 94304, USA \and
Bay Area Environmental Research Institute, NASA Research Park, Moffett Field, CA 94035, USA\\
\email{smitha@mps.mpg.de}}
\titlerunning{\ion{Ti}{I} lines at 2.2\,$\mu$m as probes of the cool parts of sunspots}
\authorrunning{Smitha et al.}
\date{Received ; accepted }

\abstract
 %  context heading (optional)
  {The sunspot umbra harbors the coolest plasma on the solar surface due to the presence of strong magnetic fields. The routinely used atomic lines to observe the photosphere have weak signals in the umbra and are often swamped by molecular lines. This makes it harder to infer the properties of the umbra, especially in the darkest regions.} 
  {The lines of the \ion{Ti}{I} multiplet at 2.2\,$\mu$m are formed mainly at temperatures $\le\!4500$\,K and are not known to be affected by molecular blends in sunspots. Since the first systematic observations in the 1990's, these lines have been seldom observed due to the instrumental challenges involved at these longer wavelengths. We revisit these lines and investigate their formation in different solar features.}
   {We synthesize the \ion{Ti}{i} multiplet using a snapshot from 3D MHD simulation of a sunspot and explore the properties of two of its lines in comparison with two commonly used iron lines at 6302.5\,\AA{} and $1.5648\,\mu$m. }
   {We find that the \ion{Ti}{i} lines have stronger signals than the \ion{Fe}{I} lines in both intensity and polarization in the sunspot umbra and in penumbral spines. They have little to no signal in the penumbral filaments and the quiet Sun, at $\mu=1$. Their strong and well-split profiles in the dark umbra are less affected by stray light. {Consequently, inside the sunspot it} is easier to invert these lines and to infer the atmospheric properties, compared to the iron lines.}
   {The Cryo-NIRSP instrument at the DKIST will provide the first ever high resolution observations in this wavelength range. In this preparatory study, we demonstrate the unique temperature and magnetic sensitivities of the Ti multiplet, by probing the Sun's coolest regions which are not favourable for the formation of other commonly used spectral lines. We thus expect such observations to advance our understanding of sunspot properties.}
   

\maketitle
%____________________________________________________%%%%%%%%

\section{Introduction}
\label{sec:intro}
A regular sunspot mainly consists of a dark umbra surrounded by penumbra with its filamentary structure. The umbra can further host relatively brighter features such as umbral dots, and in some cases light bridges. Each penumbral filament is made of a bright filament head, which is located closer to the umbra and a tail which is the radially outward end of the filament. The darker regions in between the bright filaments are called spines \citep{1993ApJ...418..928L}. For an overview on sunspot formation, structure and properties see \citet{2003A&ARv..11..153S, 2011LRSP....8....4B, 2019PASJ...71R...1H}. 

The umbra is the darkest and the coolest part of a sunspot, and has the strongest magnetic field.  Routinely used photospheric spectral lines such as the \ion{Fe}{I} 6301.5\,\AA{}, 6302.5\,\AA{}, 6173\,\AA{} in the visible or the \ion{Fe}{I}\,1.5648\,$\mu$m and 1.5652\,$\mu$m lines in the infrared have very weak intensity signals in the umbra. In addition, the low temperatures in the umbra offer ideal conditions for the formation of molecular lines resulting in heavy blending of the atomic lines in the visible and {infrared \citep[see Fig.5 of][for the blending of the \ion{Fe}{I}\,1.5648\,$\mu$m line ]{1992A&A...263..312S}}. 
This makes the inference of atmospheric properties from these atomic lines, by means of inversions of the Stokes profiles, quite challenging despite their strong polarization signals. Both these issues can be addressed using the Ti lines at 2.2\,$\mu$m. The lines in this multiplet have low excitation potential, low ionization potential and thus high temperature sensitivity. They are formed mostly in regions with temperature below 4500\,K \citep{1998A&A...338.1089R} such as the sunspot umbra, with weak to no signal in the penumbra or the quiet Sun. Some of the lines in the multiplet are unblended or mainly blended with telluric lines {that can be easily removed} \citep{1998A&A...338.1089R}. These characteristics make the Ti lines at 2.2\,$\mu$m ideal for isolated umbral observations with little contamination from the penumbra or the quiet Sun. 

The \ion{Ti}{I} multiplet at 2.2\,$\mu$m was first observed in a sunspot umbra for an infrared atlas by \citet{1973aiss.book.....H} at the Kitt Peak National Observatory, and later for the atlas by \citet{1992adsu.book.....W}.  Their diagnostic potential was first noted by \citet{1985ApJ...299L..47S}, who used them to make the first ever detection of magnetic field on an M dwarf, which also happened to be the strongest field discovered up to that point on a cool star \citep[][]{1994IAUS..154..437S}. Later, \citet{1987LIACo..27..103S} used the Ti multiplet to study late K and M dwarfs. {Recently, the Ti lines were observed by \citet{2009AIPC.1094..124K} to measure magnetic fields on M dwarfs using the cryogenic high-resolution cross-dispersed infrared echelle spectrograph  \citep[CRIRES,][]{2004SPIE.5492.1218K} on the Very Large Telescope (VLT) of the European Southern Observatory. For a historical context} on how these lines shaped our understanding of the stellar magnetic fields, see \citet{1994IAUS..154..437S, 1996IAUS..176..237S}. 

%===Table 1=====================
\begin{table*}[htbp]
\label{tab:atomicdata}
\begin{center}
\caption{Atomic details of the \ion{Ti}{I} multiplet at 2.2\,$\mu$m and two other iron lines (\textit{bottom two rows}) often used for sunspot diagnostics.}
\begin{tabular}{cccccc}
\toprule[1.5pt]
Wavelength (\AA{}) & Land\'{e} $\rm g_{\rm eff}$-factor & Ex. pot. (e.v.) & Ion. pot. (e.v.) & $\log(gf)$ & 
$\lambda \rm g_{\rm eff}$ ($\lambda$ in $\mu$m)\\
\midrule[1.5pt]
21897.38 & 1.17 & 1.74 & 6.83 & -1.39 & 2.54\\
22004.50 & 1.00 & 1.73 & 6.83 & -1.83 & 2.20\\
22211.22 & 2.00 & 1.73 & 6.83 & -1.71 & 4.61\\
22232.91 & 1.67 & 1.74 & 6.83 & -1.61 & 3.70\\
22274.07 & 1.58 & 1.75 & 6.83 & -1.71 & 3.51\\
22310.61 & 2.50 & 1.73 & 6.83 & -2.07 & 5.58\\
\midrule[1.5pt]
 6302.45 & 2.50 & 3.69 & 7.90 & -1.24 & 1.58\\
15648.52 & 3.00 & 5.43 & 7.90 & -0.68 & 4.70\\
\bottomrule[1.5pt]
\end{tabular}
\end{center}
\end{table*}
%===Table 1=====================
\begin{figure*}[htbp]
\centering
\includegraphics[width=\textwidth]{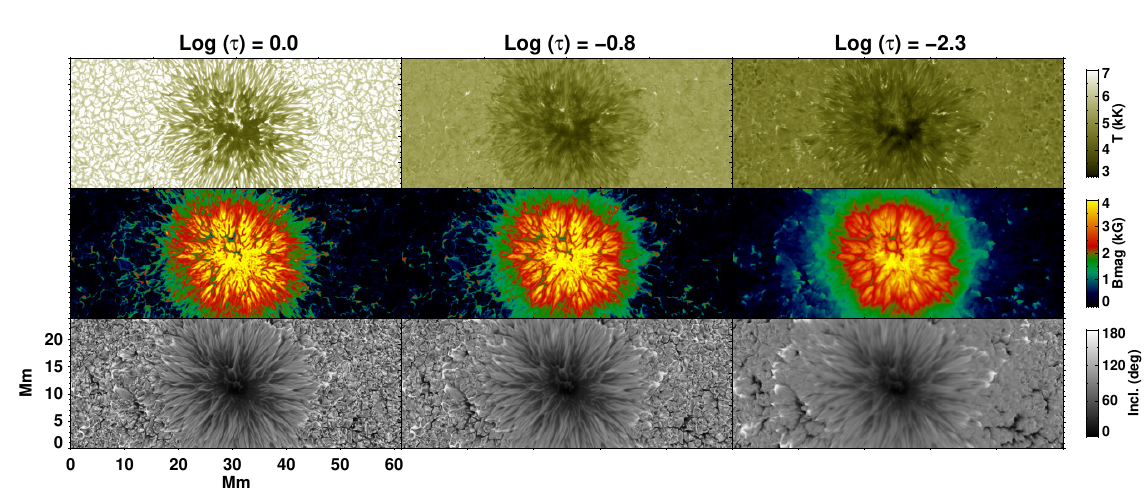}
\caption{Maps of temperature (\textit{first row}), magnetic field strength (\textit{second row}) and magnetic field inclination (\textit{third row}) from the MHD sunspot simulation at $\log(\tau)=0.0, -0.8$ and $-2.3$.}
\label{fig:mhdcube}
\end{figure*}

In the context of solar magnetic field measurements, the potential of the Ti lines was first noted by \citet{1995A&AS..113...91R}. Later \cite{1998A&A...338.1089R} presented the first systematic scans of the sunspot umbra observed using the \ion{Ti}{I} 2.2\,$\mu$m lines. Using these observations, the authors found that the sunspots are made of two distinct cool components, one  component with large vertical magnetic field associated with the umbra and another weak inclined magnetic component in the penumbra. The same observations were later used by \citet{1999A&A...348L..37R} to study the Evershed flow in the cooler channels of the penumbra. Observations in the \ion{Ti}{I} line at 2.231\,$\mu$m were used by \cite{2003SoPh..215...87P, 2003ApJ...590L.119P} to measure the velocities of the Evershed outflow, as well as umbral and penumbral magnetic fields. The same line will also be observed by LOCNES: low cost NIR extended solar telescope to study the effects of cooler plasma on the disk-integrated flux and mean magnetic field {on the Sun in order to understand the stellar jitter in radial velocity technique arising due to activity on other stars \citep{2018SPIE10700E..4NC}. The radial velocity technique is used for extrasolar planets detection.}

The high temperature sensitivity and a large magnetic sensitivity of the \ion{Ti}{I} lines form a unique combination to probe the sunspot umbra. Their Zeeman sensitivity is {larger} than visible lines such as the commonly used \ion{Fe}{I} 6300\,\AA{} pair observed by \textit{Hinode/SOT-SP} \citep{Ichimoto2008SoPh}. As discussed in \cite{1998A&A...338.1089R}, the \ion{Ti}{I} lines are less affected by molecular blends and stray light from the surrounding penumbra and quiet Sun regions. Other than the McMath–Pierce Telescope at the National Solar Observatory at Kitt Peak, no other telescopes have been used to observe these interesting lines {on the Sun}, partly due to the lack of instruments that can observe at these wavelengths. However, the Cryo-NIRSP instrument \citep{2016SPIE.9908E..4DF} available at the newly constructed DKIST \citep{2014SPIE.9147E..07E, 2020SoPh..295..172R} will be able to observe the Ti lines at $2.2\,\mu$m for the first time at a high spatial and spectral resolution. 

In the present paper, we investigate the diagnostic potential of \ion{Ti}{I} line around 2.2 $\mu$m. To this end we synthesize the Ti multiplet from a sunspot simulation of \citet{2012ApJ...750...62R} along with two commonly used iron lines in the visible and infrared. We then carry out a comparative study to explore the unique potential of the Ti lines and how they can be used to study different features of a Sunspot. 

\section{The \ion{Ti}{I} multiplet at $2.2\,\mu$m}
\label{sec:ti-lines}
The titanium multiplet of interest has six lines \citep{2006MNRAS.373.1603B, doi:10.1063/1.3656882}, and five of them  were discussed in \cite{1998A&A...338.1089R}. The atomic details of all the lines taken from the Vienna Atomic Line Database (VALD) are presented in Table~\ref{tab:atomicdata}. We have also included the two commonly observed iron lines at 6302\,\AA{} and 15648\,\AA{} in the table. 

{The Zeeman splitting of a spectral line increases as $\lambda^2$g$_{\rm eff}$, where $\lambda$ is the central wavelength and g$_{\rm eff}$ is the effective Land\'{e} factor. Since the Doppler width of the line increases linearly with $\lambda$, the observable magnetic splitting of the spectral line is, therefore, $\propto\!\lambda$g$_{\rm eff}$. For the titanium multiplet and the two iron lines, $\lambda$g$_{\rm eff}$  are indicated in the last column of Table~\ref{tab:atomicdata}. In the titanium multiplet, the 22310\,\AA{} line has the largest magnetic sensitivity which is 3.5 times larger than the \ion{Fe}{I} 6302\,\AA{} line and nearly 1.2 times the sensitivity of the \ion{Fe}{I} 1.56\,$\mu$m line}. From the Zeeman splitting patterns shown in \citet{1998A&A...338.1089R}, the \ion{Ti}{I} 22310\,\AA{} is the only line in this multiplet with a normal Zeeman triplet. But it is weaker than the other lines due to its smaller line strength ($\log(gf)$). 

{From the low-resolution observations presented in \cite{1995A&AS..113...91R}, four lines in the titanium multiplet at 22211\,\AA{}, 22232\,\AA{}, 22274\,\AA{} and 22310\,\AA{} have no known blends of solar origin. Among these, the most magnetically sensitive line {at 22310\,\AA{} with g$_{\rm eff}$=2.5} has a strong telluric blend, which, however, can be easily removed using the procedure described by the above authors and originally proposed by \cite{1973aiss.book.....H}, cf \cite{1985ApJ...299L..47S}. Another interesting line in the multiplet is the one 22211\,\AA{}, with the second largest Land\'{e} factor g$_{\rm eff}$=2.0. This line is again blended by telluric lines but they too are easily removable \citep{1995A&AS..113...91R}. The only line, that we know of, with a blend of solar origin is the one at 21897\,\AA{} which has a comparatively small g$_{\rm eff}$. Recently, high-resolution observations of the \ion{Ti}{I} 22274\,\AA{} line in M dwarfs using CRIRES by \citet[]{2009AIPC.1094..124K} revealed that it is unblended in stars hotter than M2. On the Sun, high-resolution observations from DKIST will be able to give more insights into the presence of blends close to the Ti lines.}

{For further analysis in the rest of the paper, we consider the profiles of only two lines in the titanium multiplet with extreme properties: the line with largest magnetic sensitivity and smallest $\log(gf)$ at 22310\,\AA{}, and the one with largest line strength but small Land\'{e} factor (g$_{\rm eff}=1.17$) at 21897\,\AA{}.} 

\section{Spectral profiles}
\label{sec:spectra}

\subsection{Synthesis}
\label{subsec:synthesis}
{To investigate the formation of titanium lines in a sunspot and the quiet Sun, we used a three-dimensional magnetohydrodynamic (MHD) simulation \citep{2012ApJ...750...62R, 2015ApJ...814..125R} generated using the MURaM code \citep{2005A&A...429..335V}.} The simulation box extends to  $(61.44\times61.44\times2.97)$\,Mm in the $x,y,z$ directions, respectively, with a grid spacing of  48\,km in the $xy-$direction and 24\,km in the vertical direction. {However, we only use a piece of the full cube for spectral synthesis.} The maps of temperature, magnetic field strength and magnetic field inclination from this MHD cube at $\log(\tau)=0.0,-0.8$ and $-2.3$ are shown in Figure~\ref{fig:mhdcube}. We refer to $\log(\tau_{5000})$ simply as $\log(\tau)$.

The Stokes profiles $(I,Q,U,V)$ were synthesized in local thermodynamic equilibrium (LTE) by solving the 1D radiative transfer equation along each column of the MHD cube, a scheme known as 1.5D LTE, using the SPINOR code \citet{1987PhDT.......251S, 2000A&A...358.1109F} at $\mu = 1.0$, i.e. all computations correspond to solar disc centre. As previously mentioned, we discuss the synthetic Stokes profiles of only two lines from the Ti multiplet at 21897\,\AA{} and 22310\,\AA{} and compare them with the profiles of \ion{Fe}{I} 6302.5\,\AA{} and 15648\,\AA{}. 

\begin{figure}
\centering
\includegraphics[width=0.35\textwidth]{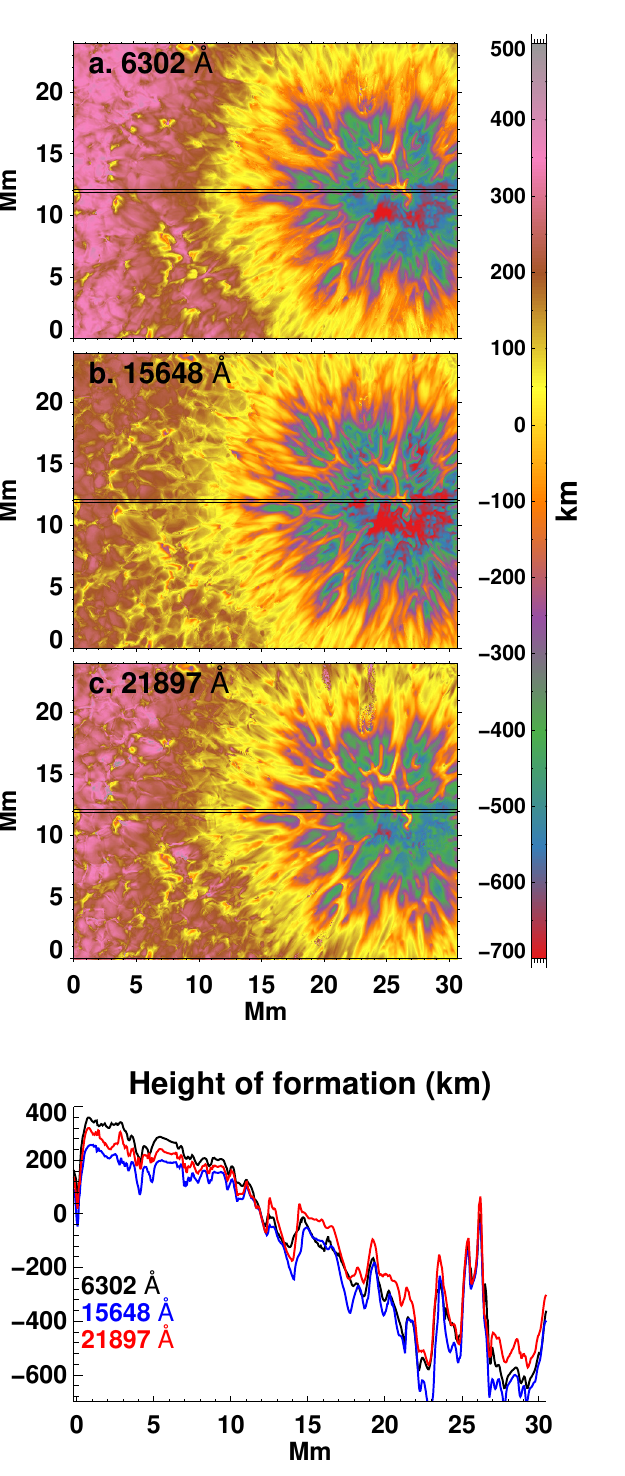}
\caption{Maps of height of formation of the three lines $\lambda$=6302.5\,\AA{} (\textit{panel a}), 15648.5\,\AA{} \textit{(panel b}), and 21897.4\,\AA{} (\textit{panel c}). The height of formation is the centroid of the intensity response function to temperature. In the bottom panel, we have plotted the heights of formation along the horizontal black lines, averaging in the $y$-direction over the narrow region between the two black lines {covering 4 pixels}.} 
\label{fig:hof}
\end{figure}

\subsection{Height of formation}
\label{subsec:hof}
To determine the atmospheric layers sampled by the titanium lines, we compute their \textit{height of formation}, in the same way as in \cite{2017A&A...608A.111S}. The height of formation of a given line is assumed to be the centroid of the range of heights sampled by the intensity profiles weighted by their response functions \citep[][]{1975SoPh...43..289B}. Here we have used the intensity response to perturbations in temperature. The maps of the height of formation so obtained are shown in Figure~\ref{fig:hof}. The heights are indicated with respect to a reference $z = 0$\,km geometrical layer which corresponds to  $\log(\tau)$=0, averaged over the simulation box. {The \ion{Ti}{I} 22310\,\AA{} line is formed at  similar heights as the 21897\,\AA{} line and hence we only discuss the latter in detail. }

The heights of formation of the three lines at \ion{Fe}{I} 6302\,\AA{}, \ion{Fe}{I} 15648\,\AA{}, and the \ion{Ti}{I} 21897\,\AA{} averaged over 4 pixels in the $y$-direction covering different features in the cube are shown in the bottom panel. The fall in the formation height of the lines from the quiet Sun to the umbra due to the Wilson depression is clearly seen. In the quiet Sun, the formation height of the titanium line falls in between the two iron lines. In the sunspot penumbra and the umbra, the titanium line is formed slightly higher than the Fe lines. The \ion{Fe}{I} infrared line at 1.56\,$\mu$m is formed the lowest everywhere in the simulation domain. These results are in agreement with the heights determined using the contribution functions by \citet{1998A&A...338.1089R}.  

\begin{figure*}
\begin{center}
\includegraphics[width=\textwidth]{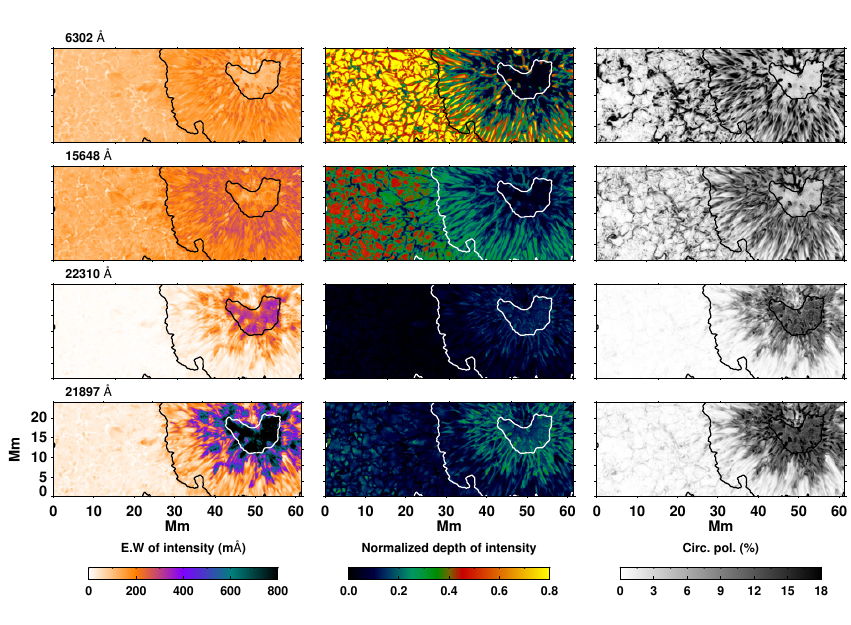}
\caption{{In the first and second columns we present maps of the equivalent widths and normalized line depths of the intensity profiles. The last columns are maps of maximum circular polarization. All quantities are plotted for four lines the \ion{Fe}{I} 6302.5\,\AA{} and 1.56\,$\mu$m lines (\textit {top panels}), and the \ion{Ti}{I} 22310\,\AA{} and 21897\,\AA{} lines (\textit{bottom panels})}. Contours mark the penumbra-umbra and penumbra-quiet sun boundaries. }
\label{fig:ew_vbyi_mu1.0}
\end{center}
\end{figure*}

 \begin{figure*}
 \begin{center}
 \includegraphics[width=\textwidth]{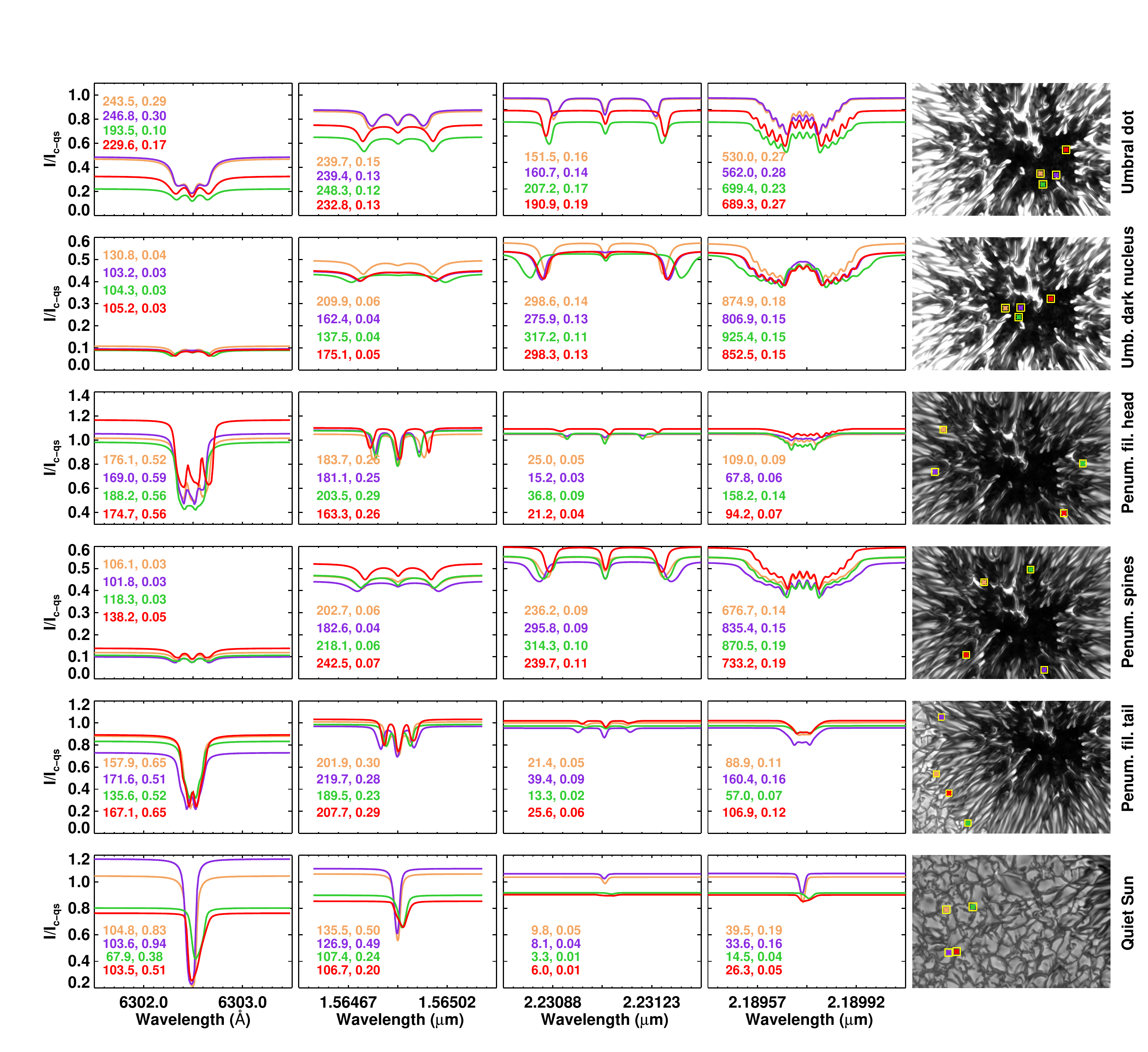} 
 \caption{Stokes $I$ profiles from different features of the sunspot and the quiet Sun. From top to bottom these are: umbral dots, dark umbral core, penumbral filament head, spine in penumbra, penumbral filament tail and quiet Sun. At each feature, we show profiles from four sample pixels {and indicate the equivalent widths (in m\AA) and normalized line depths.} These pixels are at the centers of the squares marked on the images in the last column. The profiles are shown for both the \ion{Fe}{i} and the \ion{Ti}{i} spectral lines. They are normalized to the spatially averaged quiet Sun continuum intensity ($I_{\rm c-qs}$). The contrast on the last column was adapted for better representation of the different features.}
  \label{fig:int_profiles}
  \end{center}
\end{figure*}

 \begin{figure*}
 \begin{center}
 \includegraphics[width=0.9\textwidth]{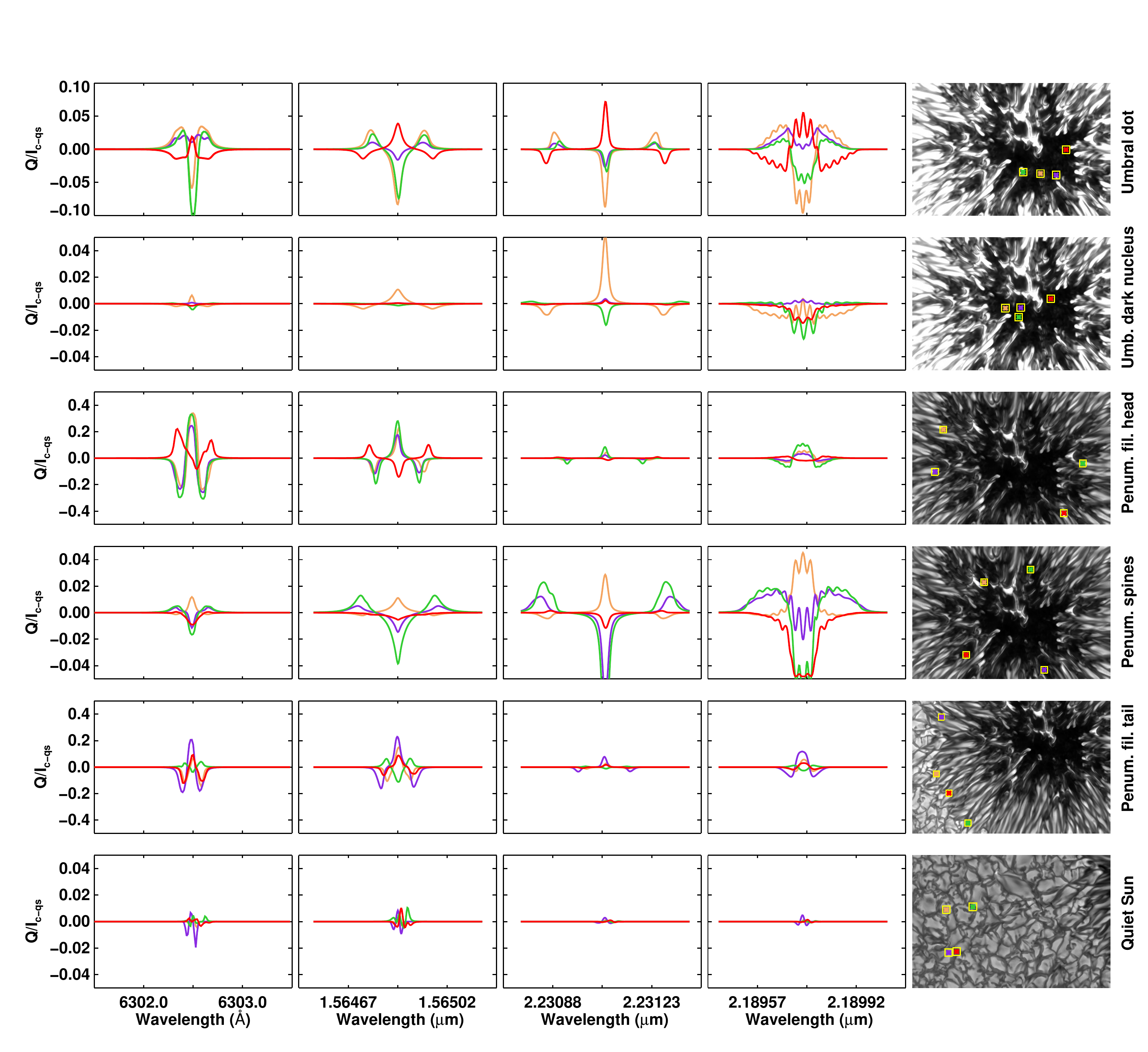} 
 \caption{Same as Figure~\ref{fig:int_profiles} but for the $Q/I_{\rm c-qs}$ profiles.} 
\label{fig:qbyi_profiles}
\end{center}
\end{figure*}

\subsection{Stokes profiles in different features}
\label{stokes-profiles}
\subsubsection{Normalization}
The synthetic Stokes profiles computed from the sunspot cube can be normalized in two different ways, one with the spatially averaged quiet Sun intensity ($I_{\rm c-qs}$) and the other using the local continuum in that pixel ($I_{\rm c-loc}$). In the literature we find both approaches being used depending on the problem that is addressed. In the brighter features such as the granules, the $I_{\rm c-qs}$ is smaller than $I_{\rm c-loc}$ and in the dark sunspot umbra, the $I_{\rm c-loc}$ is only a small fraction of $I_{\rm c-qs}$. Thus the Stokes profiles have different amplitudes depending on the normalization used. Since the normalization with $I_{\rm c-qs}$ is more commonly used for the inversion of Stokes profiles \citep{2000ApJ...534..989B, Socas_Navarro_2004, Riethm_ller_2008, 2013A&A...554A..53R,  2015A&A...583A.119T}, we follow the same in the rest of the paper.

\subsubsection{Equivalent widths and normalized line depths}

In Figure~\ref{fig:ew_vbyi_mu1.0}, we compare the equivalent widths (EW; \textit{first column}), normalized line depths (NLD, the line depth of Stokes $I$ profiles normalized to the average quiet Sun continuum intensity, \textit{middle column}), and maximum circular polarization (\textit{last column}). Maps are presented for the \ion{Ti}{I} 22310\,\AA{} and 21897\,\AA{} lines along with the two iron lines at 6302\,\AA{} and 15648\,\AA. In the quiet Sun, both EW and NLD are small for the titanium lines. For the \ion{Ti}{I} 22310\,\AA{} line, the NLD is $<0.05$ and EW is $< 10$\,m\AA{}. Due to its slightly larger $\log(gf)$, EW reaches $50$\,m\AA{} for the \ion{Ti}{I} 21897\,\AA{} line along with NLD  up to $0.2$. In comparison, the two iron lines at 6320\,\AA{} and at $1.56\,\mu$m have much larger EW and NLD in the quiet Sun. Here the EW of the \ion{Fe}{I} 6302\,\AA{} line ranges from $50$\,m\AA{} - $150$\,m\AA, with the NLD ranging from $0.2 - 1.3$. The NLD of the \ion{Fe}{I} 6302\,\AA{} line exceeding $1.0$ are commonly from pixels where the local continuum is higher than the spatially averaged quiet Sun continuum. The infrared iron line at $1.56\,\mu$m, although it has smaller NLD ($<0.55$) compared to the visible line, has larger EW due to its longer wavelength. In the quiet Sun, the EW of the iron $1.56\,\mu$m line varies between $50 - 200$\,m\AA. From the circular polarization maps in  Figure~\ref{fig:ew_vbyi_mu1.0}, both titanium lines at 21897\,\AA{} and 22310\,\AA{} have little to no $V/I_{\rm c-qs}$ signal in the quiet Sun while the $V/I_{\rm c-qs}$ of the iron lines reaches values as high as $10\% - 15\%$ in kilo-gauss magnetic elements.

{In the sunspot umbra, the titanium lines have a clear advantage. The EW of the stronger line at 21897\,\AA{} ranges from $400$\,m\AA - $1.1$\,\AA{}  and the NLD varies between $0.1 - 0.4$. The distribution of EW and NLD over all the pixels peak around 900\,m\AA{} and $0.15$, respectively. (the larger values are not visible in Figure~\ref{fig:ew_vbyi_mu1.0} as the EW color scale  is saturated at 800\,m\AA{} so that the change in EW of the other lines can be seen as well. {The 21897\,\AA{} line undergoes anomalous Zeeman splitting and we see multiple Zeeman components in the Stokes profiles \citep[see Figure 1 of][]{1998A&A...338.1089R}. Being a normal Zeeman triplet, the \ion{Ti}{I} 22310\,\AA{} line is  split into its three Zeeman components. To compute the EW, we integrate over all the Zeeman components for all the lines.} The EW and NLD in the umbra range between $100 - 400$\,m\AA{} and $0.05 - 0.2$, respectively. For both \ion{Ti}{I} 21897\,\AA{} and 22310\,\AA{} lines, the EWs measured in  \citet[see their Figs. 3 and 4]{1998A&A...338.1089R} are smaller while the NLD values are larger than in our synthesis. This is because in \citet[]{1998A&A...338.1089R} the EW and NLD were measured in the absence of a magnetic field.}

{The \ion{Fe}{I} 1.56\,$\mu$m line, due to its large Land\'e factor g$_{\rm eff}=3.0$, is fully split in the umbra and penumbra. In the umbra, its EW is comparable to the \ion{Ti}{I} 22310\,\AA{} line. The EW of the 1.56\,$\mu$m line varies between $100 - 300$\,m\AA{} with a peak of the distribution around $200$\,m\AA{}. Its NLD over the umbra is smaller than the \ion{Ti}{I} 22310\,\AA{} line and it varies between $0.02 - 0.2$ with a peak of the distribution at $0.05$. Of all the four lines presented in Figure~\ref{fig:ew_vbyi_mu1.0}, the \ion{Fe}{I} 6302\,\AA{} has the smallest EW ($70 - 250$\,m\AA{}) and NLD ($0.01 - 0.2$). }

{The circular polarization profiles of the two iron lines at 6302\,\AA{} and the 1.56\,$\mu$m are Zeeman saturated at umbral magnetic fields. The lines in the titanium multiplet including those considered in Figure~\ref{fig:ew_vbyi_mu1.0} (\textit{last column)} may seem not to suffer from Zeeman saturation, although they are clearly completely split in the umbra. This is due to a combination of low excitation potential, low ionization potential and low abundance \citep{1995A&AS..113...91R} and the anti-correlation between magnetic field strength  and temperature in sunspot umbrae \citep[e.g.][]{1990Ap&SS.170...75M, 1992SoPh..141..253K, 2015A&A...583A.119T}. In the umbra, the Ti lines at 21897\,\AA{} and 22310\,\AA{} have $V/I_{\rm c-qs}$ signals exceeding $15\%$ while for the \ion{Fe}{I} 6302\,\AA{} and 1.56\,$\mu$m lines, they do not exceed $5\%$.}

In Figures~\ref{fig:int_profiles} - \ref{fig:vbyi_profiles}, we compare the Stokes profiles, $I/I_{\rm c-qs}$, $Q/I_{\rm c-qs}$ and $V/I_{\rm c-qs}$ of the two iron lines and the two titanium lines, in pixels representing different features within the simulation. The chosen pixels are indicated on the maps by yellow boxes. We discuss them in more detail in the sections below.

\subsubsection{Umbra}
\label{umbra}
\begin{figure*}
 \begin{center}
 \includegraphics[width=0.9\textwidth]{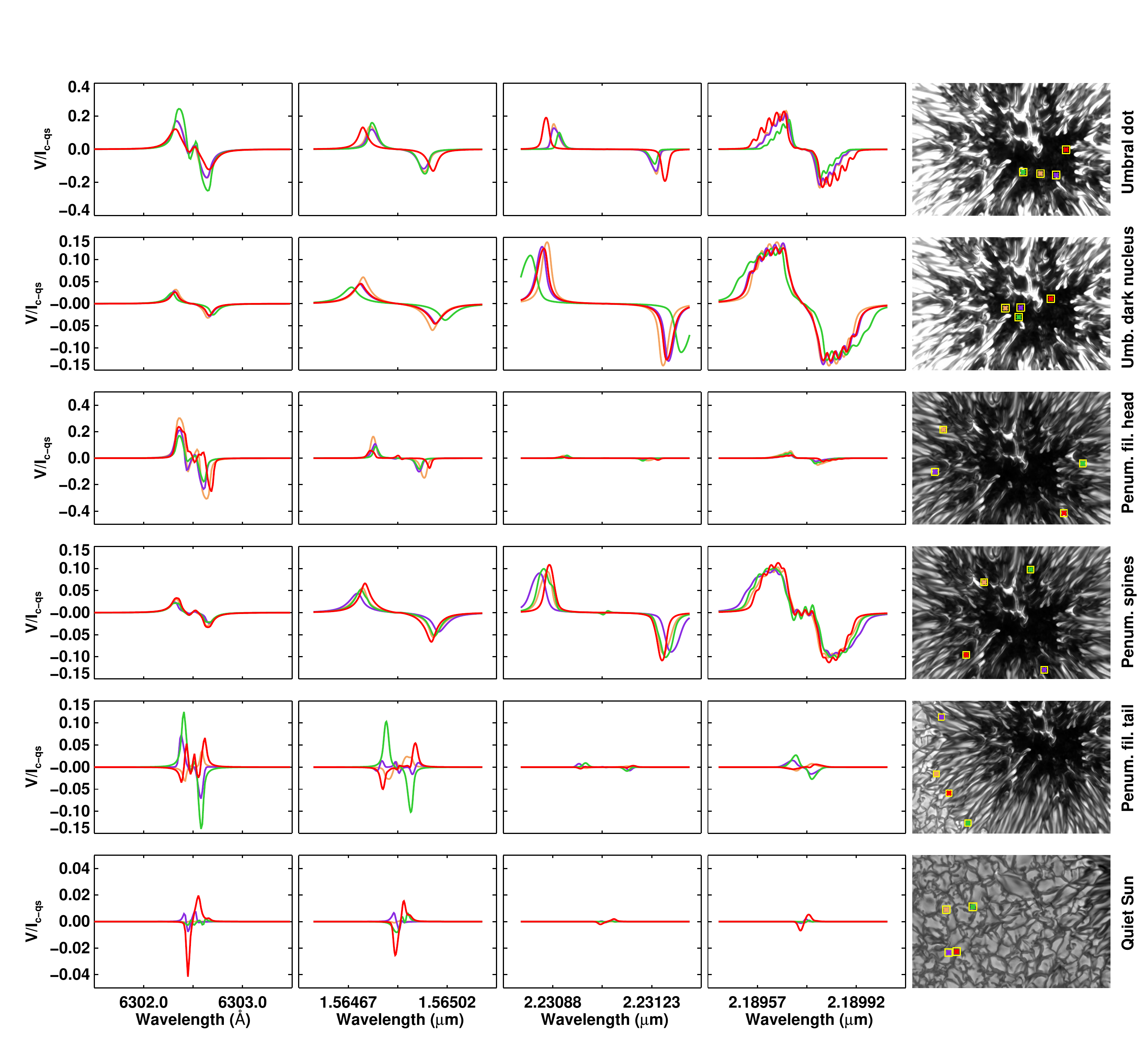} 
 \caption{Same as Figure~\ref{fig:int_profiles} but for $V/I_{\rm c-qs}$ profiles.} 
  \label{fig:vbyi_profiles}
  \end{center}
\end{figure*}

A typical sunspot umbra consists of a dark background and relatively brighter umbral dots. The dark nucleus, i.e. the part of the umbra with the darkest background and few umbral dots, which has the strongest magnetic field in the umbra \citep{1992SoPh..141..253K, 1993A&A...275..283S} generally covers 10\% to 20\% of the total umbral area \citep{2003A&ARv..11..153S}. %\sks{There are a number of papers that show that the darkest part of the umbra has the strongest field. E.g. Kopp & Rabin (1992); Solanki et al. (1993).} 
Due to the low temperature of the plasma, the conditions here are ideal for the formation of titanium lines while the iron lines have very weak signals. {In the first two columns of Figures~\ref{fig:int_profiles}-\ref{fig:vbyi_profiles}, we show intensity and polarization profiles of the two iron lines at 6302\,\AA{} and 1.56\,$\mu$m. These should be compared with the profiles of \ion{Ti}{I} 22310\,\AA{} and 21897\,\AA{} {(columns 3 and 4, respectively)}. The profiles in the first row of all three figures are from umbral dots while the second row shows examples from the dark umbral nucleus.}

{In the pixels from the umbral dots, all the four spectral lines considered in the figures, have intensity profiles with NLD greater than 0.1. {The \ion{Ti}{I} 21897\,\AA{} line has the largest NLD and EW, followed by the \ion{Fe}{I} 1.56\,$\mu$m line (see $I/I_{\rm c-qs}$ profiles plotted in the top row of Figure~\ref{fig:int_profiles}). In brighter umbral dots (\textit{indicated by purple and orange lines}), the NLD and EW of the 6302\,\AA{} intensity profiles are comparable to the \ion{Fe}{I} 1.56\,$\mu$m. In the remaining two umbral dots shown in the figure, they are smaller. On average, the \ion{Ti}{I} 22310\,\AA{} line has smaller NLD and EW in the four representative pixels from umbral dots chosen here.} The Zeeman components of the \ion{Fe}{I} 1.56\,$\mu$m and the two Ti lines are completely split in the umbral dots.  All the four lines have comparable amplitudes of both $Q/I_{\rm c-qs} (\sim 5\% - 10\%)$ and $V/I_{\rm c-qs} (\sim 5\%)$ profiles. The two iron lines have been used in several papers to study the properties of umbral dots \citep[see e.g.][]{Socas_Navarro_2004, 2008A&A...492..233R, 2009ApJ...702.1048W, 2012ApJ...757...49W, Yadav_2018}. For the reasons previously stated, observations in the Ti multiplet will provide new insights into the umbral dots and their fine-scale structures \citep{2006ApJ...641L..73S, 2007ApJ...669L..57B}. }

In the dark umbral nucleus, the \ion{Fe}{I} 6302\,\AA{} and 1.56\,$\mu$m lines nearly disappear with NLD much less than 0.1, although their EW $>100$\,m\AA{}. In comparison, the Ti lines are stronger. The EW of \ion{Ti}{I} 21897\,\AA{} is greater than 800\,m\AA{} and in a few pixels it even exceeds 1\,\AA{}. The Zeeman components of the \ion{Ti}{I} 22310\,\AA{} line are far apart and they have an EW between $250 - 350$\,m\AA{}. Both these Ti lines have NLD between $0.1 - 0.2$, which is greater than the iron lines. In polarization, $Q/I_{\rm c-qs}$ and $V/I_{\rm c-qs}$ signals in the iron lines are three times smaller in amplitude.

\subsubsection{Penumbra}
 According to the uncombed configuration of the penumbra \citep{1993A&A...275..283S, 2003A&ARv..11..153S}, the magnetic field in the sunspot penumbra is composed of two components, one is the strong and vertical component known as the spines and the other is the weak inclined component called the inter-spines or filaments \citep{1993ApJ...418..928L}. The spines are dark and cooler compared to the filaments \citep[e.g.,][]{2013A&A...557A..25T}. They offer favourable conditions for the formation of the Ti lines. From Figure~\ref{fig:int_profiles}, the EW and NLD of the two titanium lines 21897\,\AA{} and 22310\,\AA{} in the penumbral spines are larger than those of the iron lines (\textit{fourth row}). In the spines, the profiles from \ion{Fe}{I} 1.56\,$\mu$m lines are although fully split and have EW $>180$\,m\AA{}, their NLD are less than $0.1$. The EW and NLD of the \ion{Fe}{I} 6302\,\AA{} line in the penumbral spines are comparable to its profiles in the dark umbra. From Figures~\ref{fig:int_profiles}-\ref{fig:vbyi_profiles}, the two Ti lines have stronger polarization signals than the Fe lines in the dark penumbral spines. In the penumbral filament, however, the Fe lines are much stronger than the Ti lines in both intensity and polarization due to larger plasma temperatures. {\citet[]{2015A&A...583A.119T} demonstrated that penumbral filament tails are in general darker and cooler than the filament heads \citep[see also][]{2010ApJ...722L.194B, 2013A&A...557A..25T}. From rows three and five of Figures~\ref{fig:int_profiles} - \ref{fig:vbyi_profiles}, we see that the titanium lines are weak in filament head as well as in the filament tail. We do not see a significant strengthening of the signal due to drop in temperature from the head to the tail. This is because the profiles shown in the figures are selected from the tails of the outer penumbral filaments. Here the difference in temperature between the filament head and filament tail is only about 300\,K, on average \citep[][]{2013A&A...557A..25T}. Since the temperature of the tails in the inner penumbra are lower and comparable to the temperature in the spines \citep[][]{2013A&A...557A..25T}, it is possible that the Ti line has a stronger signal in the tails of inner penumbral filaments (see Figure~\ref{fig:ew_vbyi_mu1.0}). The profiles from the inner penumbral filaments are not shown in Figures~\ref{fig:int_profiles} - \ref{fig:vbyi_profiles}. Also, the tails and heads of filaments in the simulation are not as localized and distinct as in the observations -- quantitative differences between the observations and simulations do exist. }

\subsubsection{Quiet Sun}
\label{sec:quietsun}
According to Figure~\ref{fig:hof}, the Ti lines are formed in the upper photosphere. Although the granule temperature drops rapidly with height in the atmosphere, it is not cool enough for the formation of the titanium multiplet. They have nearly no signal in both intensity and polarization in the quiet Sun, with the exception of \ion{Ti}{I} 21897\,\AA{} line. Due to its large $\log(gf)$, this line has weak measurable signals, although these are not found in the immediate vicinity of the sunspot. In Figure~\ref{fig:ew_vbyi_mu1.0} (\textit{second column, bottom row}), the 21897\,\AA{} line forms a dark ring around the penumbra, signifying basically no line absorption. {This is due to the presence of the magnetic canopy, which, at the height of formation of Ti lines, extends beyond the photospheric ($\log(\tau)=0.0$) penumbral boundary  \citep{1982SoPh...79..267G, 1992A&A...263..339S, 1994A&A...283..221S}. Above the canopy base, the gas density is considerably reduced as a result of the magnetic pressure, so that the absorption in all lines is considerably reduced there. This forces much of the line to be formed below the canopy, where the temperature is relatively high. Whereas this does not affect the \ion{Fe}{I} lines very much, which are not so sensitive to temperature, the absorption due to the Ti lines is very small due to their strong temperature sensitivity. Together, this results in very weak Ti I lines.  These signals are sufficiently weak to not contaminate the sunspot observations through straylight, providing desirable conditions for studying the nature of the cool penumbral spines or the dark umbra. In the sunspot umbra, the Fe lines not only have weak signals but could also suffer contamination from the strong quiet-Sun and penumbral signals.}

\begin{figure*}
   \centering
  \includegraphics[width=.51\textwidth]{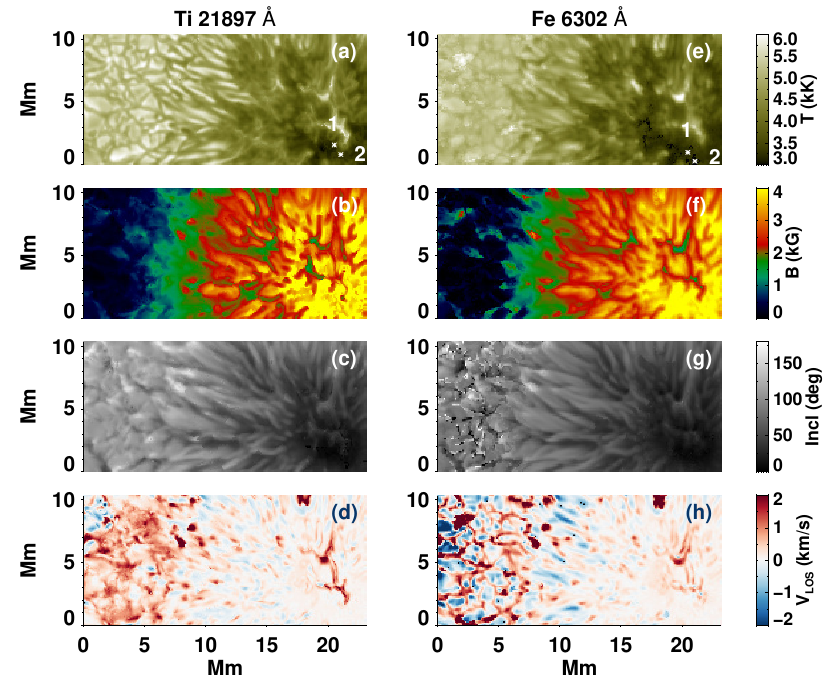}
  \includegraphics[width=.48\textwidth]{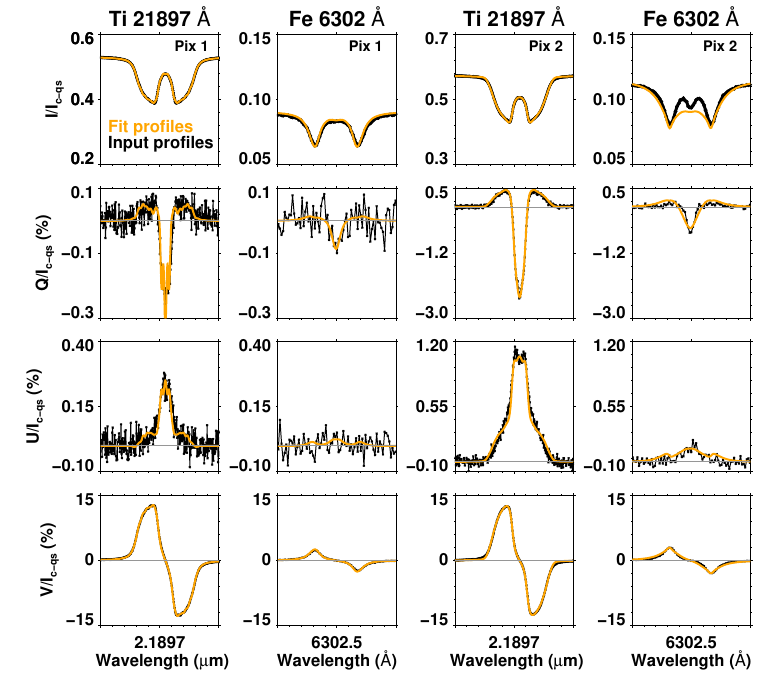}
    \caption{\textit{Left:} Maps of the temperature (\textit{a, e}), magnetic field strength (\textit{b, f}), inclination (\textit{c, g}), and LOS velocity (\textit{d, h}) at $\log(\tau)=-0.8$,  obtained by inverting the \ion{Ti}{i} 21897\,\AA{} (\textit{first column}) and \ion{Fe}{i} 6302\,\AA{} (\textit{second column}) lines. In the LOS velocity maps, blue indicates upflows and red  downflows. {\textit{Right:} Fit to the degraded Stokes profiles of the Ti and Fe lines at two sample pixels marked on the temperature maps}}
    \label{fig:invfe_map}
\end{figure*}

\section{Simulation of observations}
\subsection{Degradation}
\label{sec:degraded}
The Stokes profiles of the titanium line and the iron lines were degraded to simulate real observations. Out of the 5 lines in the titanium multiplet, we chose the one at 21897\,\AA{} since it has the largest $\log(gf)$ {and is very strong in the umbra and penumbral spines with EWs as high as $800$\,m\AA{} (see Figure~\ref{fig:int_profiles})}. This line was degraded according to the specifications of the DKIST Cryo-NIRSP instrument which has a spatial sampling of $0.15\arcsec$/pix and a spectral resolving power of 100\,000 for on-disk observations. The iron line at 6302\,\AA{} was degraded to the specifications of \textit{Hinode/SOT-SP}, which has a similar spatial sampling of $0.16\arcsec$/pix (normal mode) and a spectral resolution of $21.5$\,m\AA{}/pixel. Then a random noise was added by convolving with a Gaussian distribution with standard deviation of $1 \times 10^{-3} I_{\rm c-qs}$. The profiles were further degraded with a global $10\%$ stray light {in way similar to that described in \citet{2019A&A...630A.133M}}.

\subsection{Inversion of the Stokes profiles}
\label{sec:inv_deg}
After degradation, the profiles were inverted using the code SPINOR 1D \citep{2000A&A...358.1109F}, which is based on the STOPRO routines \citep{1987PhDT.......251S}. SPINOR solves the radiative transfer equations for polarized light to retrieve the atmospheric conditions independently at every pixel inside the field of view. We use a simple atmospheric model with 3 nodes for the temperature, magnetic field vector, LOS velocity and 1 node for microturbulent velocity. For the optimum placement of the nodes for the inversion of the \ion{Ti}{I} line, we first inverted a narrow stripe covering different features like the quiet Sun, penumbra and the darkest umbra. In each trial, we slightly shifted the nodes and calculated the mean-$\chi^2$ of the narrow stripe. A plot of the mean-$\chi^2$ as a function of the node position, gave us a parabola-like curve. The middle and top nodes were chosen based on the location where the curve reaches minimum. In the end, the nodes were placed at $\log(\tau)=0.0, -0.8$ and $-2.3$ which are also suitable for inverting the \textit{Hinode/SOT-SP} observations \citep[][]{2020ApJ...895..129C}.

{Using the single component model described above (i.e. without accounting for stray light in the inversion), we were able to easily invert the \ion{Ti}{I} 21897\,\AA{} line. However we were unable to get a good fit to the \ion{Fe}{I} 6302\,\AA{} line profiles. In order to model the stray light component in the iron line, we performed tests by adding a second component model atmosphere with 3 nodes in temperature, one each in the LOS velocity and magnetic field strength to account for any net wavelength shift due to the degradation, as well as a filling factor. {The two-component model slightly improved the fits to the iron line}. Clearly, the iron line is much more affected by stray light compared to the titanium line.} 
{In the umbra, the strong linear and circular polarization signals of the titianium line offers an additional advantage during inversions. For observations, the absence of strong solar blends is another strong advantage.} 

{In Figure~\ref{fig:invfe_map} (\textit{right panels}), we compare the fits to both titanium and iron lines using single component atmosphere at two sample pixels in the umbra. In these pixels, the \ion{Fe}{I} 6302\,\AA{} line displays a weak signal in both intensity and polarization, and is strongly affected by spatial degradation and stray light. Here the \ion{Ti}{I} 21897\,\AA{} line scores better. For testing, we inverted the Stokes profiles both with and without the degradation (Section~\ref{sec:degraded}). Our tests showed that the complexity of finding a good fit to the \ion{Fe}{I} line increases substantially after the degradation (especially in the dark regions). On the other hand, the \ion{Ti}{I} line did not show such a behaviour. The inversion retrieved good fits to the synthetic profiles in dark regions with and without the degradation.}

{In the same figure we also show maps of the inverted atmosphere obtained from the titanium and iron lines. We only show the maps at $\log(\tau)=-0.8$ since the middle node is the best constrained. Due to the difference in the height of formation of the two lines (see Figure~\ref{fig:hof}), the atmospheric maps in the fist and second column of Figure~\ref{fig:invfe_map} appear different. In the quiet Sun, the maps derived from the \ion{Ti}{i} line appear noisy due to the low signal, clearly seen in the LOS velocity map. The titanium line nevertheless provides reasonable looking results, with the quality of the fits also being good. It's strong suit is in the umbra and in penumbral spines.} 

In real observations,  heavy blending by molecular lines will further complicate the determination of atmospheric parameters in the dark umbral regions. Sample profiles from the \textit{Hinode/SOT-SP} observations illustrating the numerous weak lines blending the iron lines around 6302 \AA{} in the umbra are shown in Figure~\ref{fig:hinode_profiles} in the Appendix. {The spatially coupled inversions \citep{2012A&A...548A...5V, 2013A&A...557A..24V} might be able to fit a stray light affected profile but this remains to be tested.} On the other hand, the \ion{Ti}{I} 21897\,\AA{} line could be inverted using 1D inversion codes. {However, the caveat is that the molecular blending of the 6302\,\AA{} line only became visible in the very low stray light observations of Hinode. It therefore cannot be ruled out that such observations will reveal molecular blending also of the \ion{Ti}{I} lines} 

\section{Conclusions}
\label{sec:conclusions}

In this paper, we have explored how useful the lines in the Ti multiplet at $2.2\,\mu$m are for probing different fine-scale features of a sunspot. {Due to their low ionization potential and low excitation potential, the Ti lines are highly sensitive to temperature. They are formed only in cooler regions, increasing in strength with decreasing temperature until about 4000 K, but decreasing again below roughly 3500 K due to the formation of TiO molecules \citep[][]{1998A&A...338.1089R}. Hence they are strongest in the sunspot umbra and penumbral dark spines. Due to their longer wavelengths and large Land\'{e} factors (mainly the 22310\,\AA{} line), they are strongly magnetically sensitive ($\lambda$g$_{\rm eff}$).  All the lines in the Ti multiplet have $\lambda$g$_{\rm eff}$ greater than the commonly used \ion{Fe}{I} 6302\,\AA{}. The \ion{Ti}{I} 22310\,\AA{} line with g$_{\rm eff}=2.5$ is even somewhat more magnetically sensitive than the \ion{Fe}{I} 1.56\,$\mu$m, although this might be offset by the fact that it is formed higher in the atmosphere, where the magnetic field is generally somewhat weaker. The synthetic observations from an MHD simulation reveal that the Ti lines at 21897\,\AA{} and 22310\,\AA{}, the two out of the six lines in the multiplet chosen by us for detailed analysis, have stronger signals (larger EW, larger line depth, and stronger polarization signals) in the dark umbra and penumbral spines, compared to the \ion{Fe}{I} 6302\,\AA{} and 1.56\,$\mu$m lines. The \ion{Ti}{I} multiplet may also provide unique information about ultra cool parts of the umbra ($T<3800$\,K at $\log\tau=0$; Figure~\ref{fig:hinode_profiles}) that mostly remain unexplored.}

{Tests conducted by degrading the Stokes profiles of the \ion{Ti}{I} 21897\,\AA{} and  \ion{Fe}{I} 6302\,\AA{} lines, followed by inversions, revealed that the Ti line is much less affected by stray light compared to the iron line. {The Ti line could be more easily inverted compared to the iron line inside the simulated spot. That is, the fit to the Ti lines could be easily obtained even with a simple 3-node single component atmosphere}. Based on the observations of the Ti multiplet presented in \citet{1995A&AS..113...91R}, \citet{1998A&A...338.1089R}, \citet{2003SoPh..215...87P}, four out of six lines do not have any blend of solar origin. This is particularly advantageous for observing sunspot umbrae since the \ion{Fe}{I} 6302\,\AA{} and 1.56\,$\mu$m lines are blended by solar molecular lines.}

{Since the Ti lines sample only the regions of low temperature, they complement the Fe lines, so that both sets of lines can be used together to probe both cool and hotter structures in a sunspot. For example, the high resolution observations from \textit{Hinode} and ground-based telescopes have revealed a variety of small scale structures within the penumbra with different physical properties. The iron lines and titanium lines together can be used to understand the properties of both cool as well as hotter features within the penumbra, such as the interaction between the hot penumbral filaments and the cool spines. 

The Evershed flow in the cooler channels of the penumbra measured using the \ion{Ti}{i} line by \citet{1999A&A...348L..37R} were at a spatial resolution of $2^{\arcsec} - 3^{\arcsec}$. Till date, the Evershed flow has been studied in great detail using the current generation high resolution data at different spectral ranges. This has never been repeated in the Ti $2.2\,\mu$m lines. It will be quite interesting to how the DKIST observations will fit into our current understanding of this dynamic phenomenon. 

 {With a range of magnetic sensitivities and line strengths, the \ion{Ti}{i} multiplet at $2.2\,\mu$m offers a set of lines to choose from. The high resolution data recorded by the Cryo-NIRSP instrument on DKIST will offer many possibilities of exploiting them.}

\begin{acknowledgements}
We thank M. Rempel for kindly providing the MHD cube. HNS thanks H. P. Doerr for discussions regarding stray light. This project has received funding from the European Research Council (ERC) under the European Union's Horizon 2020 research and innovation programme (grant agreement No. 695075). J.S.C.D. was funded by the Deutscher Akademischer Austauschdienst (DAAD) and the International Max Planck Research School (IMPRS) for Solar System Science at the University of G\"ottingen. S.K.T. gratefully acknowledges support by NASA contract NNM07AA01C (Hinode).  Hinode is a Japanese mission developed and launched by ISAS/JAXA, with NAOJ as domestic partner and NASA and STFC (UK) as international partners. It is operated by these agencies in co-operation with ESA and NSC (Norway). {This work has made use of the VALD database, operated at Uppsala University, the Institute of Astronomy RAS in Moscow, and the University of Vienna.} This research has made use of NASA’s Astrophysics Data System.
\end{acknowledgements}

%\bibliography{ms}

\begin{thebibliography}{55}
\expandafter\ifx\csname natexlab\endcsname\relax\def\natexlab#1{#1}\fi

\bibitem[{{Beckers} \& {Milkey}(1975)}]{1975SoPh...43..289B}
{Beckers}, J.~M. \& {Milkey}, R.~W. 1975, \solphys, 43, 289

\bibitem[{{Bellot Rubio} {et~al.}(2000){Bellot Rubio}, {Collados}, {Ruiz Cobo},
  \& {Rodr{\'\i}guez Hidalgo}}]{2000ApJ...534..989B}
{Bellot Rubio}, L.~R., {Collados}, M., {Ruiz Cobo}, B., \& {Rodr{\'\i}guez
  Hidalgo}, I. 2000, \apj, 534, 989

\bibitem[{{Bharti} {et~al.}(2007){Bharti}, {Joshi}, \&
  {Jaaffrey}}]{2007ApJ...669L..57B}
{Bharti}, L., {Joshi}, C., \& {Jaaffrey}, S.~N.~A. 2007, \apjl, 669, L57

\bibitem[{{Bharti} {et~al.}(2010){Bharti}, {Solanki}, \&
  {Hirzberger}}]{2010ApJ...722L.194B}
{Bharti}, L., {Solanki}, S.~K., \& {Hirzberger}, J. 2010, \apjl, 722, L194

\bibitem[{{Blackwell-Whitehead} {et~al.}(2006){Blackwell-Whitehead},
  {Lundberg}, {Nave}, {Pickering}, {Jones}, {Lyubchik}, {Pavlenko}, \&
  {Viti}}]{2006MNRAS.373.1603B}
{Blackwell-Whitehead}, R.~J., {Lundberg}, H., {Nave}, G., {et~al.} 2006,
  \mnras, 373, 1603

\bibitem[{{Borrero} \& {Ichimoto}(2011)}]{2011LRSP....8....4B}
{Borrero}, J.~M. \& {Ichimoto}, K. 2011, Living Reviews in Solar Physics, 8, 4

\bibitem[{{Castellanos Dur{\'a}n} {et~al.}(2020){Castellanos Dur{\'a}n},
  {Lagg}, {Solanki}, \& {van Noort}}]{2020ApJ...895..129C}
{Castellanos Dur{\'a}n}, J.~S., {Lagg}, A., {Solanki}, S.~K., \& {van Noort},
  M. 2020, \apj, 895, 129

\bibitem[{{Claudi} {et~al.}(2018){Claudi}, {Ghedina}, {Pace}, {Gallorini}, {Di
  Giorgio}, {Liu}, {Tozzi}, {Lanza}, {Micela}, {Molinari}, {Phillips}, \&
  {Tripodo}}]{2018SPIE10700E..4NC}
{Claudi}, R., {Ghedina}, A., {Pace}, E., {et~al.} 2018, in Society of
  Photo-Optical Instrumentation Engineers (SPIE) Conference Series, Vol. 10700,
  Ground-based and Airborne Telescopes VII, ed. H.~K. {Marshall} \&
  J.~{Spyromilio}, 107004N

\bibitem[{{Elmore} {et~al.}(2014){Elmore}, {Rimmele}, {Casini}, {Hegwer},
  {Kuhn}, {Lin}, {McMullin}, {Reardon}, {Schmidt}, {Tritschler}, \&
  {W{\"o}ger}}]{2014SPIE.9147E..07E}
{Elmore}, D.~F., {Rimmele}, T., {Casini}, R., {et~al.} 2014, in Society of
  Photo-Optical Instrumentation Engineers (SPIE) Conference Series, Vol. 9147,
  Ground-based and Airborne Instrumentation for Astronomy V, ed. S.~K.
  {Ramsay}, I.~S. {McLean}, \& H.~{Takami}, 914707

\bibitem[{{Fehlmann} {et~al.}(2016){Fehlmann}, {Giebink}, {Kuhn},
  {Messersmith}, {Mickey}, {Scholl}, {James}, {Hnat}, {Schickling}, \&
  {Schickling}}]{2016SPIE.9908E..4DF}
{Fehlmann}, A., {Giebink}, C., {Kuhn}, J.~R., {et~al.} 2016, in Society of
  Photo-Optical Instrumentation Engineers (SPIE) Conference Series, Vol. 9908,
  Ground-based and Airborne Instrumentation for Astronomy VI, ed. C.~J.
  {Evans}, L.~{Simard}, \& H.~{Takami}, 99084D

\bibitem[{{Frutiger} {et~al.}(2000){Frutiger}, {Solanki}, {Fligge}, \&
  {Bruls}}]{2000A&A...358.1109F}
{Frutiger}, C., {Solanki}, S.~K., {Fligge}, M., \& {Bruls}, J.~H.~M.~J. 2000,
  \aap, 358, 1109

\bibitem[{{Giovanelli} \& {Jones}(1982)}]{1982SoPh...79..267G}
{Giovanelli}, R.~G. \& {Jones}, H.~P. 1982, \solphys, 79, 267

\bibitem[{{Hall}(1973)}]{1973aiss.book.....H}
{Hall}, D. N.~B. 1973, {An atlas of infrared spectra of the solar photosphere
  and of sunspot umbrae, in the spectral intervals 4040 cm-1-5095 cm-1; 5550
  cm-1 -6700 cm-1; 7400 cm-1 -8790 cm-1}

\bibitem[{{Hinode Review Team} {et~al.}(2019){Hinode Review Team}, {Al-Janabi},
  {Antolin}, {Baker}, {Bellot Rubio}, {Bradley}, {Brooks}, {Centeno},
  {Culhane}, {Del Zanna}, {Doschek}, {Fletcher}, {Hara}, {Harra}, {Hillier},
  {Imada}, {Klimchuk}, {Mariska}, {Pereira}, {Reeves}, {Sakao}, {Sakurai},
  {Shimizu}, {Shimojo}, {Shiota}, {Solanki}, {Sterling}, {Su}, {Suematsu},
  {Tarbell}, {Tiwari}, {Toriumi}, {Ugarte-Urra}, {Warren}, {Watanabe}, \&
  {Young}}]{2019PASJ...71R...1H}
{Hinode Review Team}, {Al-Janabi}, K., {Antolin}, P., {et~al.} 2019, \pasj, 71,
  R1

\bibitem[{{Ichimoto} {et~al.}(2008){Ichimoto}, {Lites}, {Elmore}, {Suematsu},
  {Tsuneta}, {Katsukawa}, {Shimizu}, {Shine}, {Tarbell}, {Title}, {Kiyohara},
  {Shinoda}, {Card}, {Lecinski}, {Streander}, {Nakagiri}, {Miyashita},
  {Noguchi}, {Hoffmann}, \& {Cruz}}]{Ichimoto2008SoPh}
{Ichimoto}, K., {Lites}, B., {Elmore}, D., {et~al.} 2008, \solphys, 249, 233

\bibitem[{{Kaeufl} {et~al.}(2004){Kaeufl}, {Ballester}, {Biereichel},
  {Delabre}, {Donaldson}, {Dorn}, {Fedrigo}, {Finger}, {Fischer}, {Franza},
  {Gojak}, {Huster}, {Jung}, {Lizon}, {Mehrgan}, {Meyer}, {Moorwood}, {Pirard},
  {Paufique}, {Pozna}, {Siebenmorgen}, {Silber}, {Stegmeier}, \&
  {Wegerer}}]{2004SPIE.5492.1218K}
{Kaeufl}, H.-U., {Ballester}, P., {Biereichel}, P., {et~al.} 2004, in Society
  of Photo-Optical Instrumentation Engineers (SPIE) Conference Series, Vol.
  5492, Ground-based Instrumentation for Astronomy, ed. A.~F.~M. {Moorwood} \&
  M.~{Iye}, 1218--1227

\bibitem[{{Kochukhov} {et~al.}(2009){Kochukhov}, {Heiter}, {Piskunov}, {Ryde},
  {Gustafsson}, {Bagnulo}, \& {Plez}}]{2009AIPC.1094..124K}
{Kochukhov}, O., {Heiter}, U., {Piskunov}, N., {et~al.} 2009, in American
  Institute of Physics Conference Series, Vol. 1094, 15th Cambridge Workshop on
  Cool Stars, Stellar Systems, and the Sun, ed. E.~{Stempels}, 124--129

\bibitem[{{Kopp} \& {Rabin}(1992)}]{1992SoPh..141..253K}
{Kopp}, G. \& {Rabin}, D. 1992, \solphys, 141, 253

\bibitem[{{Lites} {et~al.}(1993){Lites}, {Elmore}, {Seagraves}, \&
  {Skumanich}}]{1993ApJ...418..928L}
{Lites}, B.~W., {Elmore}, D.~F., {Seagraves}, P., \& {Skumanich}, A.~P. 1993,
  \apj, 418, 928

\bibitem[{{Martinez Pillet} \& {Vazquez}(1990)}]{1990Ap&SS.170...75M}
{Martinez Pillet}, V. \& {Vazquez}, M. 1990, \apss, 170, 75

\bibitem[{{Mili{\'c}} {et~al.}(2019){Mili{\'c}}, {Smitha}, \&
  {Lagg}}]{2019A&A...630A.133M}
{Mili{\'c}}, I., {Smitha}, H.~N., \& {Lagg}, A. 2019, \aap, 630, A133

\bibitem[{{Penn} {et~al.}(2003{\natexlab{a}}){Penn}, {Cao}, {Walton},
  {Chapman}, \& {Livingston}}]{2003SoPh..215...87P}
{Penn}, M.~J., {Cao}, W.~D., {Walton}, S.~R., {Chapman}, G.~A., \&
  {Livingston}, W. 2003{\natexlab{a}}, \solphys, 215, 87

\bibitem[{{Penn} {et~al.}(2003{\natexlab{b}}){Penn}, {Cao}, {Walton},
  {Chapman}, \& {Livingston}}]{2003ApJ...590L.119P}
{Penn}, M.~J., {Cao}, W.~D., {Walton}, S.~R., {Chapman}, G.~A., \&
  {Livingston}, W. 2003{\natexlab{b}}, \apjl, 590, L119

\bibitem[{{Rempel}(2012)}]{2012ApJ...750...62R}
{Rempel}, M. 2012, \apj, 750, 62

\bibitem[{{Rempel}(2015)}]{2015ApJ...814..125R}
{Rempel}, M. 2015, \apj, 814, 125

\bibitem[{{Riethm{\"u}ller} {et~al.}(2013){Riethm{\"u}ller}, {Solanki}, {van
  Noort}, \& {Tiwari}}]{2013A&A...554A..53R}
{Riethm{\"u}ller}, T.~L., {Solanki}, S.~K., {van Noort}, M., \& {Tiwari}, S.~K.
  2013, \aap, 554, A53

\bibitem[{{Riethm{\"u}ller} {et~al.}(2008){Riethm{\"u}ller}, {Solanki},
  {Zakharov}, \& {Gandorfer}}]{2008A&A...492..233R}
{Riethm{\"u}ller}, T.~L., {Solanki}, S.~K., {Zakharov}, V., \& {Gandorfer}, A.
  2008, \aap, 492, 233

\bibitem[{Riethmüller {et~al.}(2008)Riethmüller, Solanki, \&
  Lagg}]{Riethm_ller_2008}
Riethmüller, T.~L., Solanki, S.~K., \& Lagg, A. 2008, The Astrophysical
  Journal, 678, L157

\bibitem[{{Rimmele} {et~al.}(2020){Rimmele}, {Warner}, {Keil}, {Goode},
  {Kn{\"o}lker}, {Kuhn}, {Rosner}, {McMullin}, {Casini}, {Lin}, {W{\"o}ger},
  {von der L{\"u}he}, {Tritschler}, {Davey}, {de Wijn}, {Elmore}, {Fehlmann},
  {Harrington}, {Jaeggli}, {Rast}, {Schad}, {Schmidt}, {Mathioudakis},
  {Mickey}, {Anan}, {Beck}, {Marshall}, {Jeffers}, {Oschmann}, {Beard},
  {Berst}, {Cowan}, {Craig}, {Cross}, {Cummings}, {Donnelly}, {de Vanssay},
  {Eigenbrot}, {Ferayorni}, {Foster}, {Galapon}, {Gedrites}, {Gonzales},
  {Goodrich}, {Gregory}, {Guzman}, {Guzzo}, {Hegwer}, {Hubbard}, {Hubbard},
  {Johansson}, {Johnson}, {Liang}, {Liang}, {McQuillen}, {Mayer}, {Newman},
  {Onodera}, {Phelps}, {Puentes}, {Richards}, {Rimmele}, {Sekulic}, {Shimko},
  {Simison}, {Smith}, {Starman}, {Sueoka}, {Summers}, {Szabo}, {Szabo},
  {Wampler}, {Williams}, \& {White}}]{2020SoPh..295..172R}
{Rimmele}, T.~R., {Warner}, M., {Keil}, S.~L., {et~al.} 2020, \solphys, 295,
  172

\bibitem[{{R{\"u}edi} {et~al.}(1999){R{\"u}edi}, {Solanki}, \&
  {Keller}}]{1999A&A...348L..37R}
{R{\"u}edi}, I., {Solanki}, S.~K., \& {Keller}, C.~U. 1999, \aap, 348, L37

\bibitem[{{R\"{u}edi} {et~al.}(1998){R\"{u}edi}, {Solanki}, {Keller}, \&
  {Frutiger}}]{1998A&A...338.1089R}
{R\"{u}edi}, I., {Solanki}, S.~K., {Keller}, C.~U., \& {Frutiger}, C. 1998,
  \aap, 338, 1089

\bibitem[{{R\"{u}edi} {et~al.}(1995){R\"{u}edi}, {Solanki}, {Livingston}, \&
  {Harvey}}]{1995A&AS..113...91R}
{R\"{u}edi}, I., {Solanki}, S.~K., {Livingston}, W., \& {Harvey}, J. 1995,
  \aaps, 113, 91

\bibitem[{{Saar}(1994)}]{1994IAUS..154..437S}
{Saar}, S.~H. 1994, in Infrared Solar Physics, ed. D.~M. {Rabin}, J.~T.
  {Jefferies}, \& C.~{Lindsey}, Vol. 154, 437

\bibitem[{{Saar}(1996)}]{1996IAUS..176..237S}
{Saar}, S.~H. 1996, in Stellar Surface Structure, ed. K.~G. {Strassmeier} \&
  J.~L. {Linsky}, Vol. 176, 237

\bibitem[{{Saar} \& {Linsky}(1985)}]{1985ApJ...299L..47S}
{Saar}, S.~H. \& {Linsky}, J.~L. 1985, \apjl, 299, L47

\bibitem[{{Saar} {et~al.}(1987){Saar}, {Linsky}, \&
  {Giampapa}}]{1987LIACo..27..103S}
{Saar}, S.~H., {Linsky}, J.~L., \& {Giampapa}, M.~S. 1987, in Liege
  International Astrophysical Colloquia, Vol.~27, Liege International
  Astrophysical Colloquia, ed. J.~P. {Swings}, J.~{Collin}, \& E.~J. {Wampler},
  103--108

\bibitem[{Saloman(2012)}]{doi:10.1063/1.3656882}
Saloman, E.~B. 2012, Journal of Physical and Chemical Reference Data, 41,
  013101

\bibitem[{{Sch{\"u}ssler} \& {V{\"o}gler}(2006)}]{2006ApJ...641L..73S}
{Sch{\"u}ssler}, M. \& {V{\"o}gler}, A. 2006, \apjl, 641, L73

\bibitem[{{Smitha} \& {Solanki}(2017)}]{2017A&A...608A.111S}
{Smitha}, H.~N. \& {Solanki}, S.~K. 2017, \aap, 608, A111

\bibitem[{Socas-Navarro {et~al.}(2004)Socas-Navarro, Pillet, Sobotka, \&
  Vazquez}]{Socas_Navarro_2004}
Socas-Navarro, H., Pillet, V.~M., Sobotka, M., \& Vazquez, M. 2004, The
  Astrophysical Journal, 614, 448

\bibitem[{{Solanki}(1987)}]{1987PhDT.......251S}
{Solanki}, S.~K. 1987, PhD thesis, ETH, Z\"{u}rich

\bibitem[{{Solanki}(2003)}]{2003A&ARv..11..153S}
{Solanki}, S.~K. 2003, \aapr, 11, 153

\bibitem[{{Solanki} \& {Montavon}(1993)}]{1993A&A...275..283S}
{Solanki}, S.~K. \& {Montavon}, C.~A.~P. 1993, \aap, 275, 283

\bibitem[{{Solanki} {et~al.}(1994){Solanki}, {Montavon}, \&
  {Livingston}}]{1994A&A...283..221S}
{Solanki}, S.~K., {Montavon}, C.~A.~P., \& {Livingston}, W. 1994, \aap, 283,
  221

\bibitem[{{Solanki} {et~al.}(1992{\natexlab{a}}){Solanki}, {R\"{u}edi}, \&
  {Livingston}}]{1992A&A...263..339S}
{Solanki}, S.~K., {R\"{u}edi}, I., \& {Livingston}, W. 1992{\natexlab{a}},
  \aap, 263, 339

\bibitem[{{Solanki} {et~al.}(1992{\natexlab{b}}){Solanki}, {R\"{u}edi}, \&
  {Livingston}}]{1992A&A...263..312S}
{Solanki}, S.~K., {R\"{u}edi}, I.~K., \& {Livingston}, W. 1992{\natexlab{b}},
  \aap, 263, 312

\bibitem[{{Tiwari} {et~al.}(2013){Tiwari}, {van Noort}, {Lagg}, \&
  {Solanki}}]{2013A&A...557A..25T}
{Tiwari}, S.~K., {van Noort}, M., {Lagg}, A., \& {Solanki}, S.~K. 2013, \aap,
  557, A25

\bibitem[{{Tiwari} {et~al.}(2015){Tiwari}, {van Noort}, {Solanki}, \&
  {Lagg}}]{2015A&A...583A.119T}
{Tiwari}, S.~K., {van Noort}, M., {Solanki}, S.~K., \& {Lagg}, A. 2015, \aap,
  583, A119

\bibitem[{{van Noort}(2012)}]{2012A&A...548A...5V}
{van Noort}, M. 2012, \aap, 548, A5

\bibitem[{{van Noort} {et~al.}(2013){van Noort}, {Lagg}, {Tiwari}, \&
  {Solanki}}]{2013A&A...557A..24V}
{van Noort}, M., {Lagg}, A., {Tiwari}, S.~K., \& {Solanki}, S.~K. 2013, \aap,
  557, A24

\bibitem[{{V{\"o}gler} {et~al.}(2005){V{\"o}gler}, {Shelyag}, {Sch{\"u}ssler},
  {Cattaneo}, {Emonet}, \& {Linde}}]{2005A&A...429..335V}
{V{\"o}gler}, A., {Shelyag}, S., {Sch{\"u}ssler}, M., {et~al.} 2005, \aap, 429,
  335

\bibitem[{{Wallace} \& {Livingston}(1992)}]{1992adsu.book.....W}
{Wallace}, L. \& {Livingston}, W.~C. 1992, {An atlas of a dark sunspot umbral
  spectrum from 1970 to 8640 cm(-1) (1.16 to 5.1 [microns])}

\bibitem[{{Watanabe} {et~al.}(2012){Watanabe}, {Bellot Rubio}, {de la Cruz
  Rodr{\'\i}guez}, \& {Rouppe van der Voort}}]{2012ApJ...757...49W}
{Watanabe}, H., {Bellot Rubio}, L.~R., {de la Cruz Rodr{\'\i}guez}, J., \&
  {Rouppe van der Voort}, L. 2012, \apj, 757, 49

\bibitem[{{Watanabe} {et~al.}(2009){Watanabe}, {Kitai}, \&
  {Ichimoto}}]{2009ApJ...702.1048W}
{Watanabe}, H., {Kitai}, R., \& {Ichimoto}, K. 2009, \apj, 702, 1048

\bibitem[{Yadav {et~al.}(2018)Yadav, Louis, \& Mathew}]{Yadav_2018}
Yadav, R., Louis, R.~E., \& Mathew, S.~K. 2018, The Astrophysical Journal, 855,
  8

\end{thebibliography}

\begin{appendix}
\section{Profiles of \ion{Fe}{I} 6301.5\,\AA{} and 6302.5\,\AA{} from the \textit{Hinode/SOT-SP} observations}
\begin{figure*}
    \centering
    \includegraphics[width=.7\textwidth]{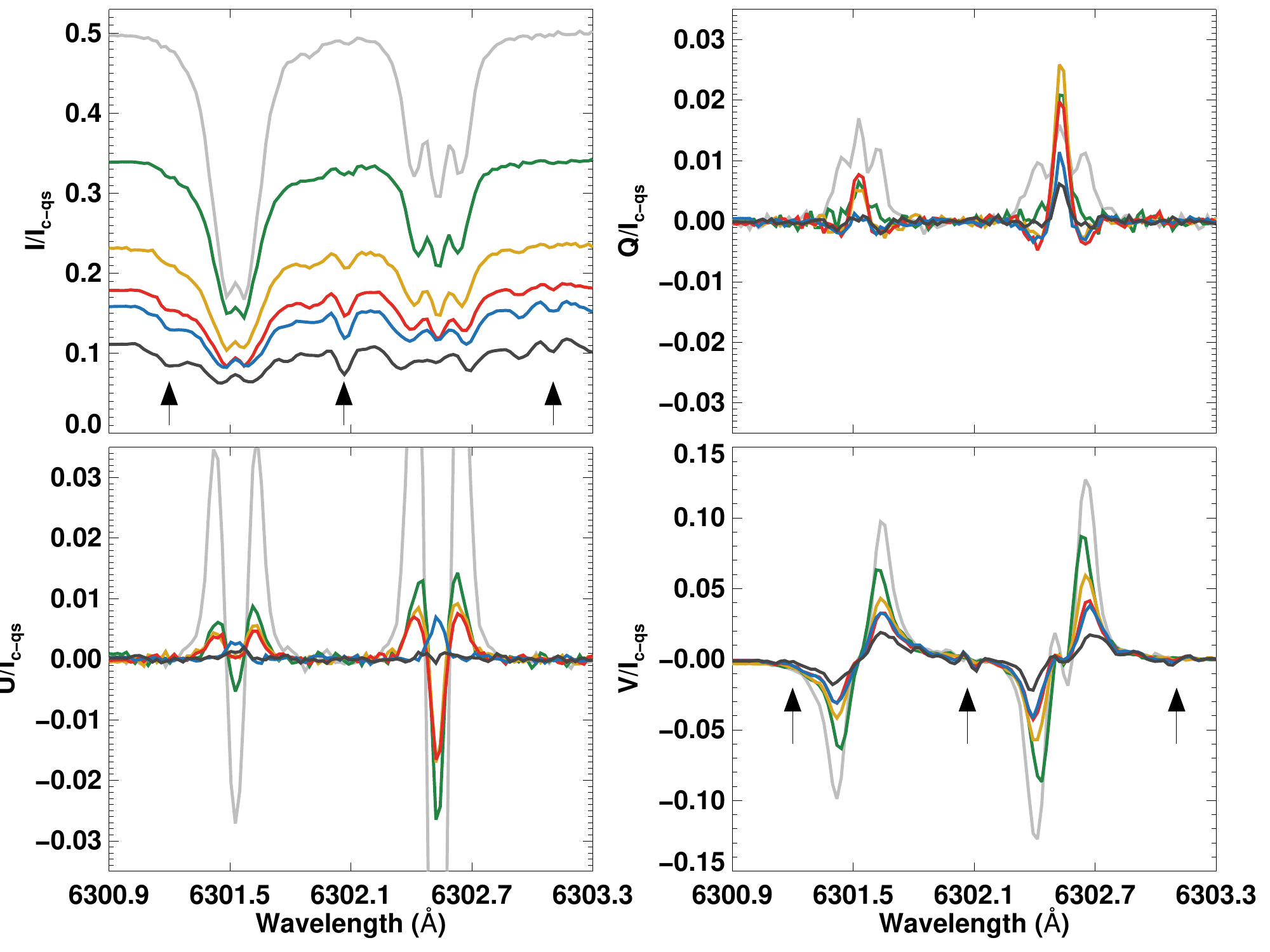}
    \includegraphics[width=.29\textwidth]{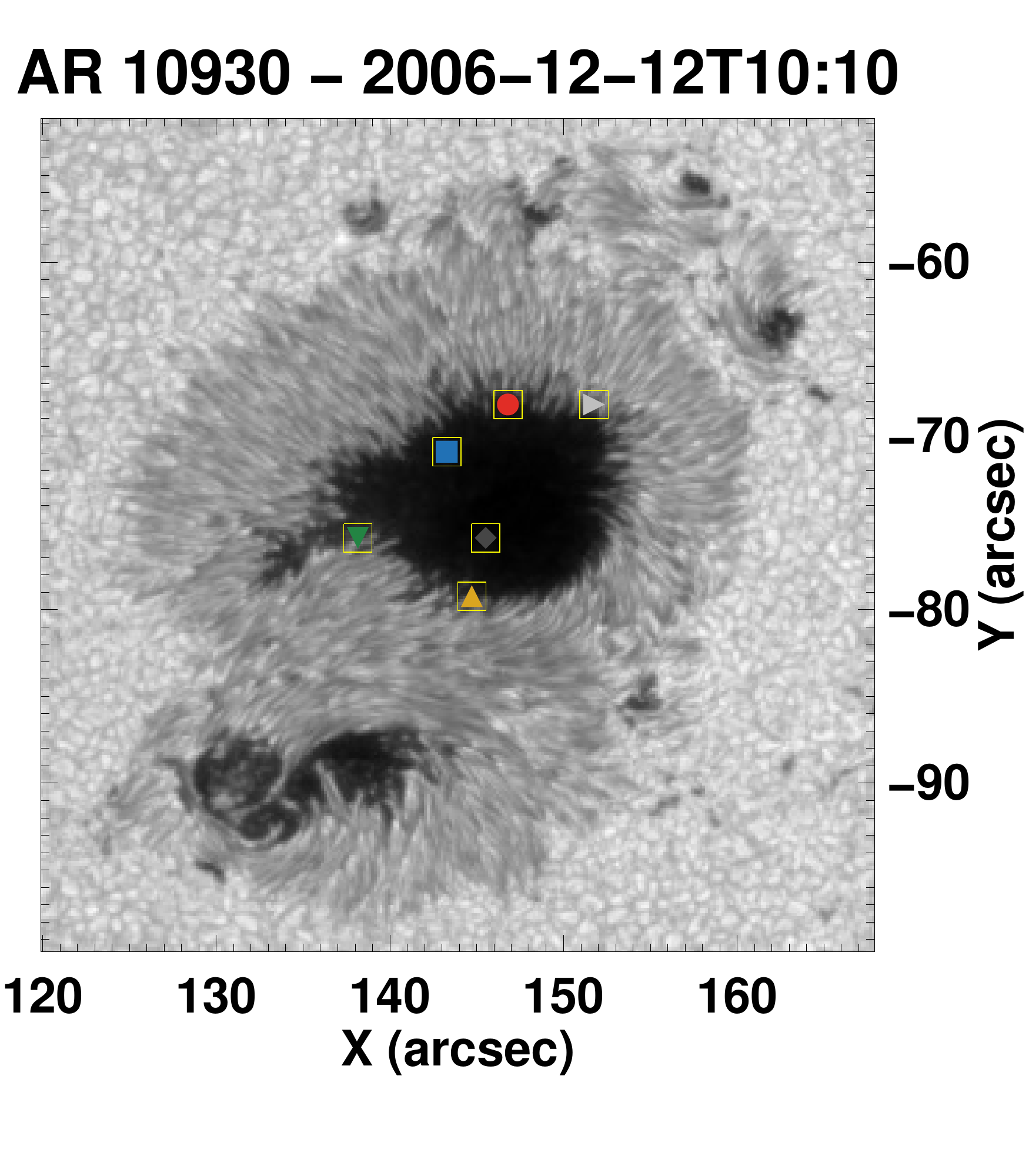}
        \caption{\textit{Hinode/SOT-SP} observations at $\mu=0.99$ of the sunspot AR\,10930. Its image at continuum wavelength is shown on the right. We have chosen six pixels, marked using squares on the image, representing different parts of the Sunspot and plotted their Stokes profiles. They sample regions with continuum intensities ranging from the darkest umbra to 0.5\,$I_{\rm qs}$.  The profiles clearly show how the 6301.5\,\AA{} and the 6302.5\,\AA{} lines become increasingly weaker with decreasing continuum intensity, while at the same time blending by most probably molecular lines increases rapidly. Arrows mark three prominent molecular lines.} 
    \label{fig:hinode_profiles}
\end{figure*} 
\end{appendix}

\end{document}